\documentclass[aps,prl,showpacs,twocolumn,superscriptaddress]{revtex4-1}

\usepackage{latexsym}
\usepackage{color}
\usepackage{bm}
\usepackage{epsfig}
\usepackage{amsfonts}
\usepackage{amssymb}
\usepackage{amsmath}
\usepackage{graphicx}
\usepackage{euscript,tabularx}
\usepackage{textcomp}
\usepackage{longtable}
\usepackage{filecontents}

\begin{document}

\title{Increased magnetic damping of a single domain wall and adjacent magnetic domains detected by spin torque diode in a nanostripe}

\author{Steven Lequeux}
\author{Joao Sampaio}
\author{Paolo Bortolotti}
\affiliation{Unit\'e Mixte de Physique CNRS/Thales and Universit\'e Paris-Sud 11, 1 Ave.~A.~Fresnel, 91767 Palaiseau, France.}

\author{Thibaut Devolder}
\affiliation{Institut d'Electronique Fondamentale, univ. Paris-Sud, CNRS UMR 8622, b\^at. 220, 91405 Orsay Cedex France.}

\author{Rie Matsumoto}
\author{Kay Yakushiji}
\author{Hitoshi Kubota}
\author{Akio Fukushima}
\author{Shinji Yuasa}
\affiliation{National Institute of Advanced Industrial Science and Technology (AIST), 1-1-1 Umezono, Tsukuba, Ibaraki 305-8568, Japan.}

\author{Kazumasa  Nishimura}
\author{Yoshinori  Nagamine}
\author{Koji~ Tsunekawa}
\affiliation{Process Development Center, Canon ANELVA Corporation, Kurigi 2-5-1, Asao, Kawasaki, Kanagawa 215-8550, Japan.}

\author{Vincent Cros}
\affiliation{Unit\'e Mixte de Physique CNRS/Thales and Universit\'e Paris-Sud 11, 1 Ave.~A.~Fresnel, 91767 Palaiseau, France.}

\author{Julie Grollier}
\affiliation{Unit\'e Mixte de Physique CNRS/Thales and Universit\'e Paris-Sud 11, 1 Ave.~A.~Fresnel, 91767 Palaiseau, France.}

\begin{abstract}
We use spin-torque resonance to probe simultaneously and separately the dynamics of a magnetic domain wall and of magnetic domains in a nanostripe magnetic tunnel junction. Thanks to the large associated resistance variations we are able to analyze quantitatively the resonant properties of these single nanoscale magnetic objects. In particular, we find that the magnetic damping of both domains and domain walls is doubled compared to the damping value of their host magnetic layer. We estimate the contributions to damping arising from dipolar couplings between the different layers in the junction and from the intralayer spin pumping effect. We find that they cannot explain the large damping enhancement that we observe. We conclude that the measured increased damping is intrinsic to large amplitudes excitations of spatially localized modes or solitons such as vibrating or propagating domain walls.
\end{abstract}

\maketitle

The spin torque diode effect provides an efficient tool to access the resonant properties of sub-micrometer magneto-resistive structures \cite {Tulapurkar}. In particular, contrarily to conventional ferromagnetic resonance techniques, its sensitivity allows probing the dynamics of individual magnetic solitons, such as vortices or domain walls, which are foreseen as information vectors in next generation magnetic devices \cite {Bedau2007, Bedau2008, Kim, Boone,Peter1,Julie,Alex1,Alex2}. By injecting a microwave current through the stack at resonance, spin torque can induce large magnetization precession of the free layer. Magneto-resistance then converts this precession into resistance variations. These resistance oscillations, multiplied by the microwave current oscillating at the same frequency, give rise to a rectified dc voltage. Most spin torque diode studies have focused on the dynamical properties of uniform magnetization configurations \cite {Sankey, Fuchs, WangJAP2009, Ishibashi,  Matsumoto} and  few on the vibration modes of magnetic solitons \cite {Bedau2007, Bedau2008, Kim, Boone,Peter1,Julie,Alex1}. Moreover, the latter studies were performed with metallic samples, where the magneto-resistance ratios are typically restricted to a few \%. These low magneto-resistance ratios strongly limit the amplitude of the output dc signal. In the case of domain wall (DW) vibrations, generally confined and of limited amplitude, it is therefore very difficult to extract other quantitative parameters than the resonance frequency of the wall \cite {Bedau2007, Bedau2008, Kim, Boone,Peter1,Julie,Alex1}.

\begin{figure}
   %\centering
	\includegraphics[width=0.5\textwidth]{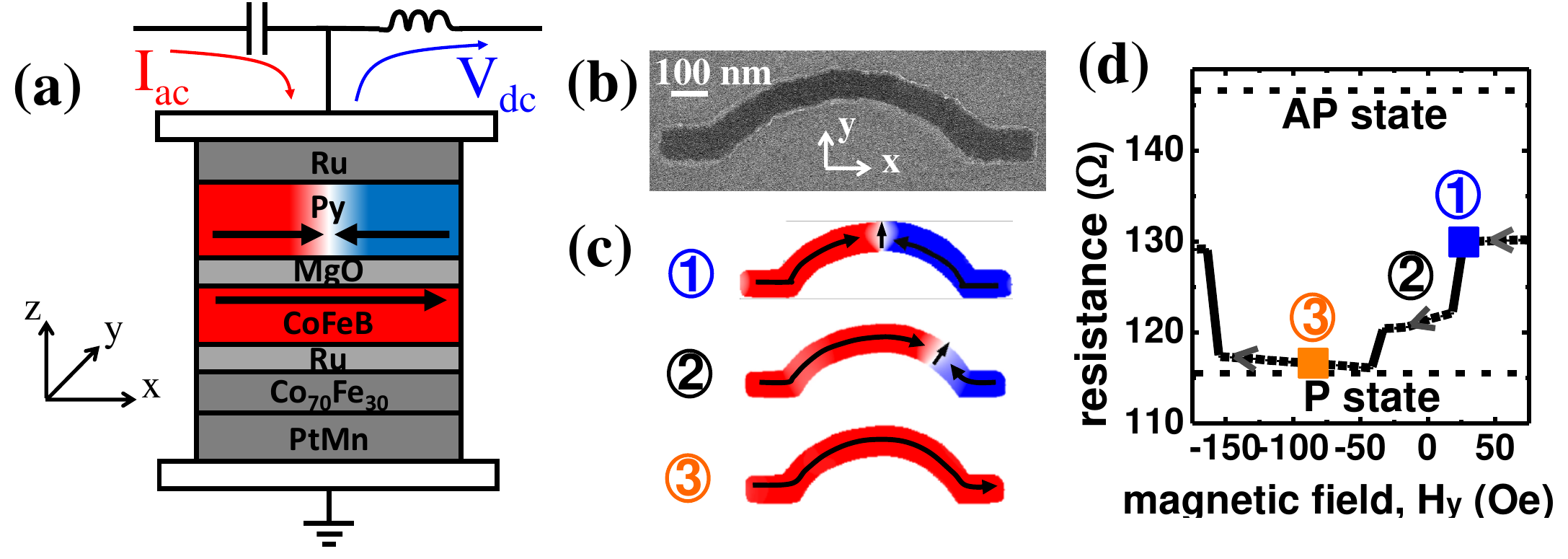}
     \caption{(a) Schematic of the spin torque diode measurement set-up including the MgO-based magnetic tunnel junction with the DW in the Permalloy free layer. (b) Scanning electron microscope image of the sample. (c) Full-stack micromagnetic simulations representing the different steps in the DW displacement. (d) Resistance versus magnetic field measurements. Black dashed lines: resistance values of the P and AP states. Black line: evolution of the resistance when a DW is nucleated (e.g. blue square at 26 Oe) in the middle of the wire (1) then pinned in the right edge (2) until its expulsion (e.g. orange square at -85 Oe) (3).}
\label{fig1}
\end{figure}

In this Letter, we take advantage of the large tunnel magneto-resistance ratios provided by MgO-based magnetic tunnel junctions to probe the dynamics of a DW and its neighboring domains by spin torque diode effect. The junctions stack, illustrated in Fig. \ref{fig1}(a), SiO$_2$//buffer/PtMn (15)/Co$_{70}$Fe$_{30}$ (2.5)/Ru (0.9)/CoFeB (3)/MgO (1.1)/Ni$_{80}$Fe$_{20}$ (5)/ Ru (10) (thicknesses in nanometers), gives rise to tunnel magneto-resistance ratios of about 30\%. A synthetic antiferromagnet (SAF) prevents the formation of a domain wall in the CoFeB reference layer. The injection of a DW in the free layer is facilitated by the arc shape geometry of our junctions, shown in Fig. \ref{fig1}(b). After the saturation of the free layer magnetization in the y direction with a strong external field (300 Oe), the DW is nucleated when the field is decreased below 75 Oe. In Fig. \ref{fig1}(c), micromagnetic simulations of the full stack realized using the OOMMF software\cite{OOMMF} show that the DW is first nucleated in the center of the wire, then displaced along the wire when the external field is decreased due to the stray field from the SAF. This scenario is in very good agreement with the experimental measurement of the junction resistance as a function of the decreasing field applied along the y direction shown in Fig. \ref{fig1}(d). The bottom and top black dashed lines, given for reference, correspond respectively to the resistance of the parallel (P) and anti-parallel (AP) states. Each change of the resistance value corresponds to a displacement of a DW in the Permalloy free layer. The first resistance plateau of 130 $\Omega$ for the field range between 75 and 26 Oe (blue square dot in Fig. \ref{fig1}(d)) corresponds to the magnetic configuration labeled 1 in Fig. \ref{fig1}(c). For lower fields, between 20 and -35 Oe, the DW shifts to a strong and reproducible pinning site due to the shape at the edge of the wire \cite{Joao}. The resulting resistance plateau at 122 $\Omega$ in Fig. \ref{fig1}(d) corresponds to the magnetic configuration labeled 2 in Fig. \ref{fig1}(c). For negative fields larger than -40 Oe, the DW is expelled of the wire and the resistance value reaches the P state at 116 $\Omega$ (orange square dot labeled 3 in Fig. \ref{fig1}(d), and magnetic configuration labeled 3 in Fig. \ref{fig1}(c)).

We probe the resonant properties of our system by spin torque diode in the configurations with and without a domain wall (respectively blue square labeled 1 and orange square labeled 3 in Fig.\ref{fig1} (b)). We inject a microwave current I$_{hf}$ between the top and bottom electrodes, and sweep the frequency. We repeat this measurement for different values of applied field H$_y$. In the absence of dc current, the rectified voltage V$_{dc}$ has two main components \cite {WangPRB2009}:

\begin{equation}
V_{dc}=\frac{1}{2} \frac{\partial^2 V}{\partial I^2} \left\langle I_{hf}(t)^2 \right\rangle + \frac{\partial^2 V}{\partial I \partial \theta} \left\langle I_{hf}(t) \theta (t) \right\rangle
\label{eqVdc}
\end{equation}

The first term is a purely electrical background signal due to the bias dependence of the tunnel junction resistance, from which we extract the exact value of the injected microwave current in the junction. The second term is the spin torque diode contribution to the rectified voltage and arises from magnetization oscillations. The parameter $\theta$ depends on the excited mode. It is related to the precession angle for quasi-uniform magnetization oscillations of the domains, and to the domain wall position for DW vibrations in a pinning potential.

\begin{figure}
   %\centering
	\includegraphics[width=0.5\textwidth]{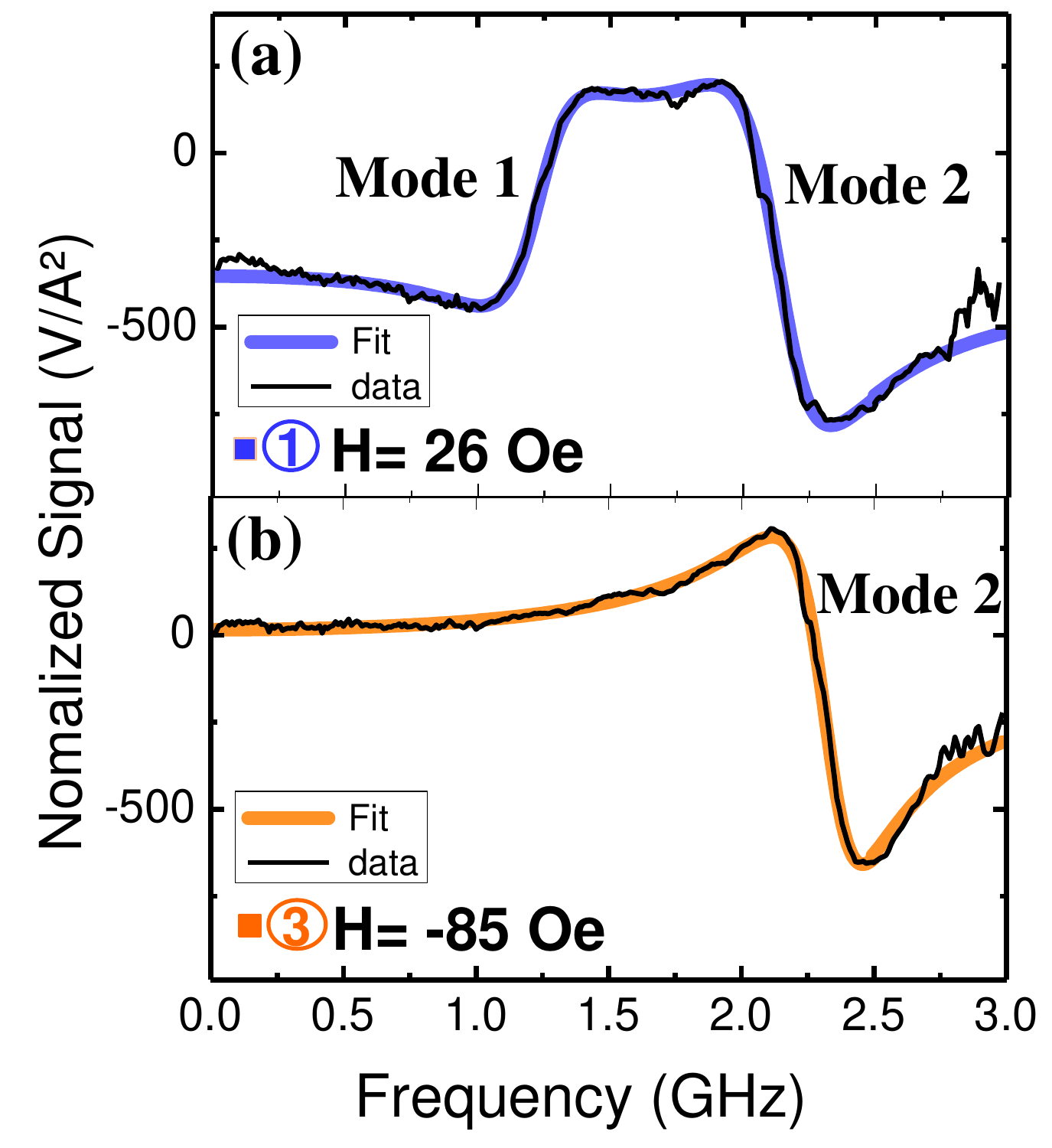}
     \caption{Normalized rectified voltage V$_{dc}$/I$_{hf}^2$ as a function of frequency (a) Black line: measurement at 26 Oe with the DW positioned in the middle of the wire corresponding to a resistance of 130 $\Omega$. Blue line: fit with Eq.(\ref{eq1}).  (b) Black line: measured at -85 Oe in the parallel state (DW expelled). Orange line: fit with Eq. (\ref{eq1}).}
\label{fig2}
\end{figure}

Fig. \ref{fig2}(a) and (b) show the rectified voltage response normalized by the square of the microwave current amplitude; V$_{dc}$/I$_{hf}^2$, at two different magnetic field values. Fig. \ref{fig2}(a) corresponds to a measurement at 26 Oe, performed at 130 $\Omega$ (state '1' in Fig. \ref{fig1}(d)), where the DW is positioned in the middle of the wire. Fig. \ref{fig2}(b) shows a measurement at -85 Oe, where the magnetization of the free layer has a quasi-uniform configuration corresponding to the parallel state (state '3' in Fig. \ref{fig1}(d)). The 26 Oe response curve shows two resonance signals at 1.24 GHz (mode 1) and 2.15 GHz (mode 2) while the -85 Oe response shows only one at 2.3 GHz (mode 2). We systematically observed two resonance signals in the field range between 50 and -35 Oe  where the DW is present whereas only the higher frequency mode (mode 2) was observed between -40 and -150 Oe where the DW is expelled. The low frequency signal (mode 1) can therefore be ascribed to DW oscillations while the higher frequency signal can be attributed to magnetization precessions in the domains. When the DW is not present mode 2 is not strongly modified, only its resonance frequency is shifted by the external field applied to allow the DW expulsion. This means that the strongest vibrations of this mode are spatially located far enough from the DW not to be impacted by the DW vibrations when the DW is still present.

\begin{figure}
   %\centering
	\includegraphics[width=0.5\textwidth]{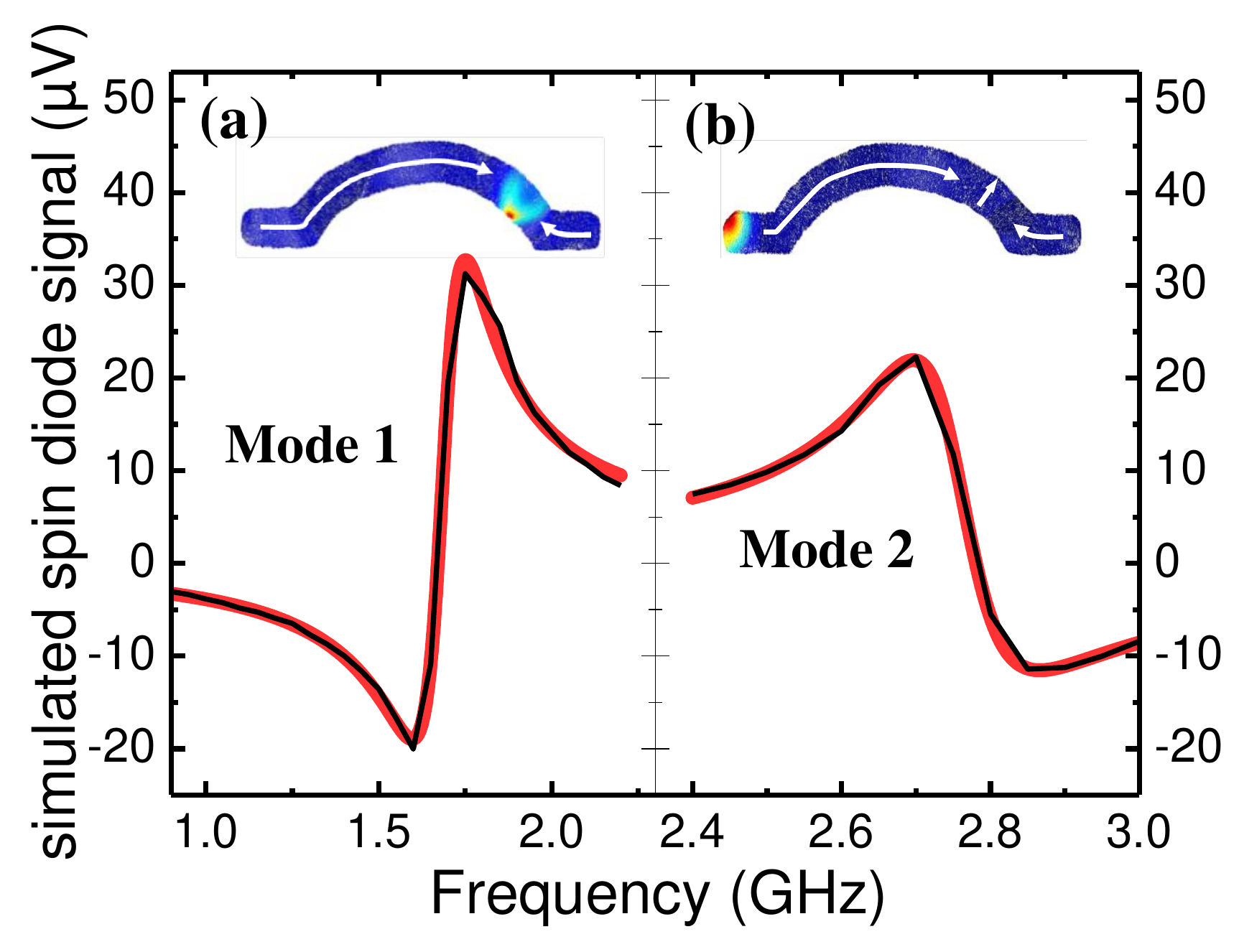}
    \caption{Spin diode signal reconstructed from micromagnetic simulations of the Py free layer at zero external field. (a) Black line: Reconstructed spin diode signal in the 0.9-2.2 GHz frequency range, showing the resonant response of the DW (mode 1). Red line: fit with Eq. (\ref{eq1}). (b) Black line: Reconstructed spin diode signal in the 2.4-3 GHz frequency range showing the domains resonant response (mode 2). Red line: fit with Eq. (\ref{eq1}). Insets: Spatial distribution of the amplitude of the magnetization oscillations dynamic component from the SpinFlow3D mode solver associated with the two observed modes. The color scale blue-green-yellow-red corresponds to the amplitude of the mode. The white arrows show the direction of the magnetization.}
\label{fig3}
\end{figure}

Around the resonance frequency \textit{f}$_{0}$, the normalized response curve takes the shape of a linear combination of Lorentzian and anti-Lorentzian profiles \cite {Tulapurkar, Matsumoto}. We fit the resonant signals of the mode 1 (DW) and 2 (domains) by the following expression:

\begin{equation}
\frac{V_{dc}}{I_{hf}^2} = \frac{A (f_0^2 - f^2) + B f^2}{(f_0^2 - f^2)^2 + (\Delta f)^2} + C
\label{eq1}
\end{equation}

where \textit{f} is the frequency of the microwave current I$_{hf}$ and the free parameters of the fit are the amplitudes A and B of the anti-Lorentzian and Lorentzian profiles, a constant shift C, the resonance frequency \textit{f}$_{0}$, and the linewidth $\Delta$. The average value of $\Delta$ extracted from the measurements by fitting with Eq. (\ref {eq1}) (Fig. \ref{fig2}) at different magnetic fields is about 0.4 $\pm$ 0.02 GHz both for mode 1 (DW) and 2 (domains). The magnetic damping parameter $\alpha$ is related to the linewidth \cite{Fuchs,WangJAP2009}. Here, in order to establish the correlation between $\Delta$ and the corresponding damping $\alpha$ for the DW and domains resonant signals we perform micromagnetic simulations of the 5 nm Py free layer at zero external field using the OOMMF software \cite {simu_monolayer,OOMMF}. We find a resonant response for the DW and domains respectively at 1.7 (Fig. \ref{fig3}(a)) and 2.75 GHz (Fig. \ref{fig3}(b)) in good agreement with the experimental results. To reconstruct the spin diode signal we sweep the frequency of the microwave current I$_{hf}$ between 0.9 and 3 GHz and extract the resulting magnetization oscillations. We plot as a function of the frequency the average amplitude of the product between the x component of the magnetization oscillations $M_{x}$ and I$_{hf}$= I$_{0}$.sin(2$\pi ft+\phi$) (Fig. \ref{fig3}). The $\phi$ parameter (respectively equal to $\frac{\pi}{2}$ and 0 in Figs. \ref{fig3}(a) and (b)) is related to the resonant signal shape depending on the symmetry of the exciting force (Slonczewski or Field like torque \cite{Tulapurkar}) and does not influence the linewidth. For a damping parameter $\alpha$=0.01, the linewidth extracted from the simulated spin diode signal by fitting with Eq.\ref {eq1} is 0.15 $\pm$ 0.002 GHz both for mode 1 and 2 (Fig. \ref{fig3}).

For the particular case of the DW resonance, in a 1D model, the linewidth $\Delta$ is related to the damping parameter $\alpha$ by:

\begin{equation}
\Delta = \frac{\gamma_{0}  \alpha  H_{d}}{2 \pi}
\label{eq2}
\end{equation}

where $\gamma_{0}$ is the gyromagnetic ratio, $\Delta$ the linewidth in Hz, and $H_{d}$ the demagnetizing field, that we calculate to be equal to 0.45 MA/m in our samples where the magnetization of the free NiFe layer is 0.47 MA/m. Based on this 1D model (Eq. \ref{eq2}), the expected linewidth from a damping of 0.01 is 0.15, in good agreement with the linewidth extracted from micromagnetic simulations. This result indicates that the 1D model is relevant to evaluate the damping extracted from the experimental linewidth measurements for the DW. Moreover, the fact that the 1D model also allows to describe the domains damping hints once again to the fact that the magnetization oscillations in the domains are spatially localized and confined. This is confirmed by the spatial distribution of the modes calculated using the mode solver from the SpinFlow3D package \cite{Peter,Spinflow_mode} where the full stack is considered \cite{full_stack_simu}, shown in the insets of Fig. \ref{fig3}(a) and (b). Indeed, the most important dynamic component of magnetization in the domains is confined and located in the left edge of the wire (inset in Fig. \ref{fig3}(b)). This edge mode is associated to the resonant response of the mode 2. For the mode 1, the resonant response is associated to a translational vibration mode of the DW (inset in Fig. \ref{fig3}(a)) \cite{Peter}.\\

Based on the previous micromagnetic simulations and the 1D model expression (Eq. \ref{eq2}), the linewidth extracted from the measurements corresponds to a damping $\alpha$ of 0.026 $\pm$ 0.001 both for DW and domains magnetization oscillations. This value of $\alpha$ is more than twice the value that we have measured by FMR on the unpatterned film (0.01$\pm$ 0.0015) and that is typically observed in Py films (0.01 \cite {Fuchs}) (see supplementary material). For each of the five samples we have measured we have obtained similar large damping values for the vibration modes, in the range 0.019 to 0.028. \\

Several mechanisms can be involved in this large increase of the effective damping compared to classical FMR measurements on thin films. We consider possible effects at stake and estimate their relative contributions. Surface roughnesses or edge defects are often pointed out \cite{Ozatay} but it has been recently shown by micromagnetic simulations that the resulting damping is an order of magnitude smaller than what we observe \cite{Silva,Weindler}.

\begin{figure}
%\centering
	\includegraphics[width=0.5\textwidth]{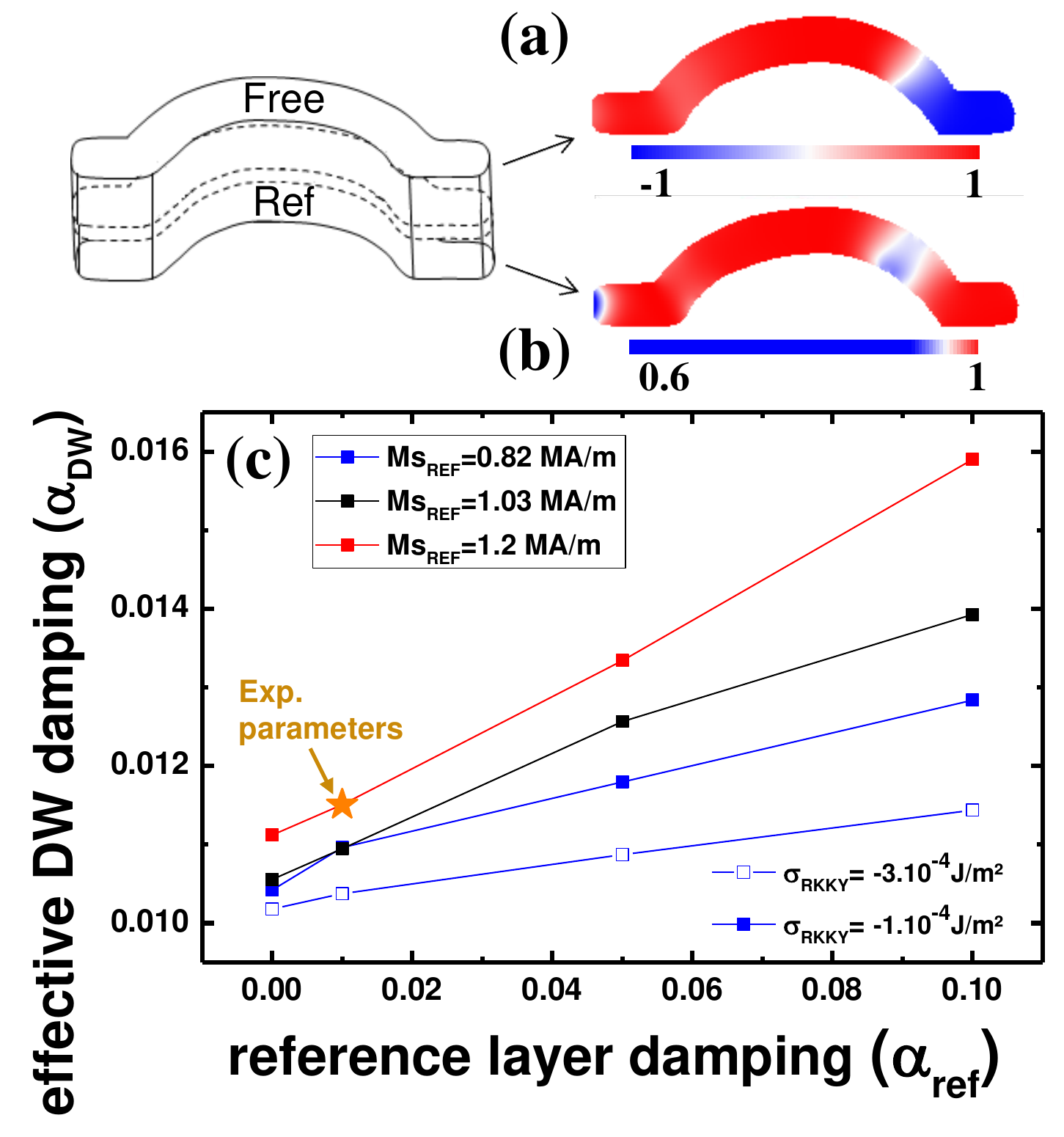}
    \caption{OOMMF micromagnetic simulations of the Permalloy free layer with the DW (a) and the CoFeB reference layer (b). (c) Dependence of the DW effective damping on the reference layers damping. The filled (resp. open) symbols correspond to an RKKY exchange coupling in the SAF of $-1.10^{-4} J/m^{2}$ (resp. $-3.10^{-4} J/m^{2}$). The value of the CoFeB reference layer magnetization is varied between 0.82 MA/m, 1.03 MA/m and 1.2 MA/m. Inset: The black solid line is the $M_{x}$ component of the NiFe layer, describing the damped oscillations of the DW. The red solid line is the fit (see supplementary material).}
\label{fig4}
\end{figure}

Another potential source of strong damping increase, especially in the case of a DW, has been pointed out by Ndjaka et al. \cite {Ndjaka}. Indeed, in trilayer structures, the stray field generated by DWs in one ferromagnetic layer dipolarly couples to the other ferromagnetic layer, leading, in the case of very strong couplings, to DW and domain duplication \cite {Thomas, Lacour}. In the case of weak coupling only a local perturbation of the magnetization in the reference layer is observed \cite{Ndjaka}. This is what we expect in our magnetic tunnel junctions as the SAF prevents domains duplication. This magnetic shadow in the reference layer coupled with the DW dynamic in the upper one results in an increased damping. To confirm and quantify this behavior, we have performed micromagnetic simulations of the full stack \cite{full_stack_simu} using the OOMMF software \cite {OOMMF}. Figs. \ref{fig4}(a) and (b) show respectively the magnetization of the Py free and the CoFeB reference layer. As expected, we observe that the magnetization of the reference layer reveals a non-uniformity located just below the DW of the free layer. We extract the effective DW damping $\alpha_{DW}$ from the DW damped oscillations (inset of Fig. \ref{fig4}(c)) as a function of $\alpha_{ref}$, the damping of the SAF layers (Fig. \ref{fig4}(c)) (see supplementary material).

In the simulations, we set the damping of the free layer to 0.01, and vary the damping of the SAF layers between 0 and 0.1. Figure \ref{fig4} (c) shows the dependence of $\alpha_{DW}$ on $\alpha_{ref}$ for two values of the Ruderman-Kittel-Kasuya-Yosida (RKKY) interlayer exchange coupling in the SAF: $-1.10^{-4}$ (filled symbols) and $-3.10^{-4}$ (open symbols) J/m$^2$ and for three values of the reference CoFeB layer magnetization: 0.82 MA/m, 1.03 MA/m and 1.2 MA/m. In each case, we observe an increase of $\alpha_{DW}$ with $\alpha_{ref}$, thus confirming the dynamic coupling between the DW and its magnetic shadow. However, in our experiments, the typical values of the RKKY exchange coupling, $\alpha_{ref}$ and Ms for the CoFeB reference layer are respectively -0.1 erg/cm$^2$, 0.01 and 1.2 MA/m \cite{Devolder,Ikeda}. As shown in Figure \ref{fig4} (c), the corresponding $\alpha_{DW}$ is about 0.0115, meaning that the DW damping increase due to dipolar coupling is one order of magnitude smaller than what we observe.

An alternative phenomena that can lead to strong damping increases is intralayer transverse spin pumping \cite {Weindler, Silva, Zhang, Tserkovnyak}. In a conducting ferromagnet, the conduction electrons carry away the excess angular momentum of a precessing magnetization. For spatially inhomogeneous magnetization dynamics, a nonuniform spin current is induced, resulting in a spatial dependance of the dissipative flow of conduction electrons in the magnetic layer itself. This can give rise to an enhanced damping in isolated magnetic nanostructures. This intralayer spin current results in the following additional nonlocal torque in the Landau-Lifshitz-Gilbert equation \cite{Silva}: $\sigma_{T} ( \overrightarrow{m} \times \nabla^{2} \frac{\partial \overrightarrow{m}}{\partial t})$, where $\sigma_{T}$ is the transverse spin conductivity. In Ref. \cite {Zhang}, the increase of damping due to intralayer spin pumping is estimated by $\eta (\frac{2 \pi}{\lambda})^{2}$, where $\eta = \frac{g \mu_{B} \hbar G_{0}}{4 e^{2} M_{S}}$, is a material dependent parameter related to $\sigma_{T}$ of Ref. \cite{Silva}. The parameter $\lambda$ is the wavelength for the magnetization pattern we consider. This means that the more magnetization dynamics is spatially located the smaller is $\lambda$, and the more the increase of damping is important. The spatial profile of the calculated modes shown in inset of Fig. \ref{fig3} can give us a rough estimate for $\lambda$, typically 100 nm for both DW and domains. By taking the parameter $\eta$ equal to 0.76 nm$^{2}$ for Permalloy with $M_{S}$=0.47 MA/m and the conductivity $G_{0}=(5 \mu \Omega   cm)^{-1}$ \cite{Zhang}, we find that the associated damping increase is 0.003. Recent analytical calculations confirm that for transverse domain walls damping increases due to intralayer spin pumping are very small \cite{JVKcondmat}. \\

Finally, spin transfer torque can give rise to large amplitude motion of magnetic objects. In our experiments, the injected microwave current density is 1.43 10$^5$ A/cm$^2$, about 10 times smaller than the critical current densities needed to depin the domain wall \cite{Joao}. From the measured rectified voltages in Fig. \ref{fig2} we find that the amplitude of domain wall translation at resonance is 4 nm, a value far from negligible compared to the domain wall width of 100 nm. It has been shown experimentally that non-linear contributions, that are not accounted for in micromagnetic simulations, can give rise to very large damping increases in nanostructures \cite{Boone2}. By elimination it seems that only these non-linear contributions appearing for wide vibration amplitudes can be responsible for the very large damping values that we measure. Our results are in phase with the high damping values generally derived from magnetic field or spin torque driven domain wall motion over large distances \cite{BoulleReview}. They also show that the damping of simple objects like transverse domain wall in a standard material such as Permalloy are still far from being understood.

As a conclusion, we have shown that by combining the spin diode effect with the large magneto-resistance ratios provided by magnetic tunnel junctions, we can quantitatively analyze the resonance of individual magnetic confined modes.  The effective damping of the domain wall and edge modes extracted from the rectified spin diode signal is more than twice increased compared to the one extracted from FMR measurements on the extended stack. We ascribe our observations to non-linear contributions to the damping appearing for large amplitudes of vibration. Our results underline that damping mechanisms are still far from being elucidated in simple systems. They also show that non-linear contributions should be taken into account when modeling large amplitude vibrations or translations of solitons in magnetic nanostructures, in particular for memory and logic applications.

\subsection*{Acknowledgements}
The authors acknowledge financial support from the European Research Council (Starting Independent Researcher Grant No.~ERC 2010 Stg 259068) and the French Ministry of Defense (DGA).


\begin{thebibliography}{1}
\bibitem{simu_monolayer}
The 5 nm Py free layer is concidered with a cell size of 4.4 nm x 4.4 nm x 5 nm (where the shape corresponds to the SEM image of Fig. \ref{fig1}). The magnetic parameters are: Ms=0.47 MA/m, $\alpha$=0.01 and A = 13 pJ/m and the spin transfer torque parameters are the current polarization of 0.3 and out-of-plane to in-plane torque ratio of 0.1.
\bibitem{full_stack_simu}
The full magnetic stack and associated parameters considered for micromagnetic simulations are: CoFe pinned [$M_s$=0.69 MA/m] (2.5 nm)/Ru (1 nm)/CoFeB [$M_s$=1 MA/m](3 nm)/MgO (1 nm)/NiFe free [$M_s$=0.47 MA/m] (5 nm).
\end{thebibliography}
\end{document}